\documentstyle[12pt]{article}  

\def\R{ {\rm R \kern -.31cm I \kern .15cm}}
\def\C{ {\rm C \kern -.15cm \vrule width.5pt \kern .12cm}}
\def\Z{ {\rm Z \kern -.27cm \angle \kern .02cm}}
\def\N{ {\rm N \kern -.26cm \vrule width.4pt \kern .10cm}}
\def\1{{\rm 1\mskip-4.5mu l} }
\def\lsim{\raise0.3ex\hbox{$<$\kern-0.75em\raise-1.1ex\hbox{$\sim$}}}
\def\gsim{\raise0.3ex\hbox{$>$\kern-0.75em\raise-1.1ex\hbox{$\sim$}}}
\def\noi{\noindent}

\def\beq{\begin{equation}}   \def\eeq{\end{equation}}
\def\bea{\begin{eqnarray}}  \def\eea{\end{eqnarray}}
\def\nn{\nonumber}
\def\noi{\noindent}
\def\beeq{\begin{eqnarray}} \def\eeeq{\end{eqnarray}}

\begin{document} 
\vbox to 1 truecm {}
\centerline{\large\bf Possible derivation of wave function reduction}\par \vskip 3 truemm
\centerline{\large\bf  from the basic quantum principles}

\vskip 1 truecm

\centerline{\bf Roland Omn\`es\footnote{e-mail: Roland.Omnes@th.u-psud.fr}}
\centerline{Laboratoire de Physique Th\'eorique\footnote{Unit\'e Mixte de
Recherche (CNRS) UMR 8627}}  \centerline{Universit\'e de Paris XI, B\^atiment
210, F-91405 Orsay Cedex, France}

\vskip 2 truecm

\begin{abstract}
Reduction is shown to be a possible consequence of the basic principles of quantum mechanics, involving no branching of the quantum state of the universe. The key feature of a measurement is attributed to the creation of macroscopic germs of ionized tracks, bubbles, sparks and so on, in a superposed quantum state, the main mechanism being a breaking and a restoration of ``classicality'' in nearby macroscopic objects and their small patterns. A new significance of quantum probabilities, involving no absolute randomness, is also proposed.
 \end{abstract}

\vskip 1 truecm
\noi PACS codes: 03.65 w, 03.65 Ta, 03.65 Yz
 
\vskip 3 truecm
\noindent LPT Orsay 08-25 \par
\noindent February 2008

\newpage
\pagestyle{plain}
\baselineskip 18pt
\section{Introduction}
\hspace*{\parindent}
The basic axioms of quantum mechanics state essentially that: {\it (i)} The possible states of a physical system belong to a Hilbert space, {\it i.e.} they satisfy the superposition principle and a scalar product implying a norm is defined on their set. {\it (ii)} Their time evolution is unitary. Three other axioms belong to interpretation, namely: {\it (iii)} The possible final states of an experimental device measuring an observable are associated with the eigenvalues of this observable. {\it (iv)} The actual final state of the measuring device is unique. {\it (v)} The occurrence of actual results in a series of identical measurements is random and the corresponding probabilities are given by Born's rule.\par
 
The first three axioms are formal whereas Axioms {\it (iv)} and {\it (v)} deal with actual experiments, Axiom {\it (iv)} states basically the uniqueness of physical reality and is often expressed explicitly as a reduction of the state of the measured system. Many consequences of the basic axioms have been developed in the last few decades, including decoherence, which removed macroscopic interferences (see for instance [1-3]), consistent histories, which disposed of logical paradoxes [4], and derivations of classical physics [1, 5], which showed particularly that classical determinism is a special outcome of quantum probabilism [5]. As for the relations between these axioms, the present situation can be summarized as follows: Axiom {\it (iii)} is a consequence of {\it (i)} and {\it (ii)} (see for instance [5]). Zurek has also shown recently that {\it (v)} follows from these basic principles [6], if {\it (iv)} is understood in the sense of Everett's ``many-worlds'' interpretation, which implyies a strong fundamental difference between the notions of reality and actuality [7]. \par
 
The present work tries to go farther with a scheme that could allow a derivation of {\it (iv)}-{\it (v)} from the basic principles {\it (i)}-{\it (ii)}, axiom {\it (iii)} being taken as already established. The scope is limited however to a statement of some convenient concepts, with no precise proof or detailed mechanism. The outcome can only stand therefore as a conjecture or as the definition of a research program.\par

Contrary to previous approaches relying on decoherence [1-3], there is no emphasis on ``pointers'' or ``meters'' registering the results of experiments, since they are only a man-made convenience providing an observer with some information, which is certainly not essential  [8]. Among the effects of decoherence, the main one is not supposed to be a selection of stable channels during a measurement process [2], but a preservation of classicality [1], ``classicality'' being understood as the collection of properties characterizing a macroscopic object obeying almost exactly classical physics. \par
 
Some new ideas are also introduced. One of them consists in characterizing a measurement process by the creation of mesoscopic and macroscopic ``germs'', such as ionization tracks leading ultimately to sparks, bubbles and so on, with or without a pointer. All the possible germs are in a quantum superposition so that they do not show a classical behavior, even when they become macroscopic, this distinction between ``macroscopic'' and ``classical'' going back to Leggett  [9]. It will be shown that non-classical germs break down the classical behavior of many objects in their vicinity, this breaking extending all around rapidly so that the first step in reduction is this contagious breaking of classicality. A convenient conceptual framework for understanding the consequences of this effect is not then Zurek's Axiom zero [6], stating that the universe is made of systems obeying quantum mechanics, but a recognition of the present universe as unique vast system, organized by a ``skeleton'' consisting of many classical objects, which themselves impose restrictions and boundary conditions on the ``tissue'' of particles and quantum fields all around them. The existence of this organization is due to two features resulting from a more precise statement of axioms {\it (i)} and {\it (ii)}, namely the existence of macroscopic bound states and the action of gravitation favoring the actual creation of such states. Finally, it will be shown that the last step in reduction can be a return to classicality according to seminal earlier ideas by Pearle [10], from which one will understand why the final state of a measuring device is unique and random at the end of an individual measurement whereas Born's rule agrees with the probabilities of the set of results in a series of so-called identical measurements. These probabilities will be associated however with the erratic fluctuations of some squared norms ({\it i.e.} of usual quantum ``probabilities'') and not with standard quantum randomness, so that one may say that ``God does not play dice'' in this theory.\par
 
Reduction would look then as a creation of non-classical macroscopic germs, bringing a breaking of classicality all around, with a final and global restoration of classicality. Some properties of classicality are therefore worth stressing. Classical interactions dominate almost everything in usual conditions, in solids as well as fluids, electric and magnetic fields, and radiation, down to a scale of a few tens of angstroms for the smallest parts in solids. These small and big features, including the objects and their smallest parts will be called ``patterns'' in the following and they are the place where classicality is broken and restored. They are usually described by collective observables (Lagrange coordinates in a classical version) and their existence is empirically obvious. Their construction from the basic quantum principles is more involved however and few works have been devoted to it (see however [11]). There is certainly a hierarchy among them, according to the size of corrections to classicality involving $\hbar$ when finer details are included, but this point is apparently inessential for the present purpose. \par
 
Classical interactions dominate the dynamics of patterns and share some specific properties, not present in the ``environment'' consisting of the atoms, particles and photons obeying the boundary conditions resulting from the classical skeleton. These classical interactions are strong, they often act altogether at a distance (as when one thinks of wheelwork or electric forces for instance) and, moreover, they are many-body forces, {\it i.e.} everything interacts with everything, directly or indirectly. The laws of classical physics describing their dynamics have been derived from the principles of quantum mechanics, especially Axiom {\it  ii}, but this derivation relied on a strong condition on the initial state of the collective subsystem (the skeleton), which was supposed to satisfy a ``classical property'' specifying simultaneously definite ranges of values for the collective observables and the related momenta [5]. In other words, classical dynamics is valid as an excellent approximation {\it when classicality is initially established}. It may be noticed that  there are exceptions to classicality (for instance when there exist very narrow potential barriers or when chaos takes place and reaches very small scales), but these cases can be considered as inessential.

\section{Reduction versus decoherence }
\hspace*{\parindent}
One will consider as an example the measurement of an observable $Z$ with eigenvalues $z$, $z'$, ... , a charged particle going along different trajectories according to these values. Geiger counters $D$, $D'$, ... , are located along these trajectories. Let one consider first the simplest case when $Z$ has only one eigenvalue and there is a unique detector $D$ whose response is usually considered as certain. The number of possible germs is certainly larger than the number of atoms in the dielectric inside $D$, since every atom can be ionized directly by the incoming particles and many different atoms are ionized along every possible trajectory of this particle. The possible initial germs are aligned along different possible tracks [12], because of quantum interferences between different Feynman histories. The germs become rapidly mesoscopic and macroscopic though their tracks are still in a state of quantum superposition and one will denote by an index $j$ these various possible tracks, differing by their location and the repartition of germs along them. The number of these indices is much larger than the number of atoms in the dielectric, so that the usual assumptions of decoherence theory do not apply: one cannot separate a fixed collective system from an environment. Although the result of the measurement is certain and a pointer would show a unique result, a unique track is selected finally, (or at least there is a unique final spark). Even the simplest measurement is not therefore completely described by decoherence theory (or at least this theory should be considerably refined) and reduction differs from decoherence. 

\section{Evolving quantum probabilities }
\hspace*{\parindent}
Decoherence theory considered only distinct results for the measurement, with an index $m$ taking as many values as the number of $Z$-eigenvalues and a collective system consisting usually of a pointer, given once and for all, without creation of germs. The non-diagonal elements $\rho_{mn}$, $m \not= n$, of the collective density matrix (reduced density) vanished exponentially with time whereas the diagonal elements remained constant (and given by Born's rule). \par

This time invariance of the diagonal elements does not hold however when the existence of germs is taken into account.  To see why, it will be convenient to introduce a set of coarse-grained parallelepipeds, with a small basis and a long side, parallel to the direction of the incoming particle. Such a region will be still called a track, although now its main features (particularly the number of electrons and ions in it) are macroscopic after a sufficient number of secondary ionizations. Another index, still denoted by $j$, will be used to distinguish these coarse-grained tracks, the number of these indices being now finite. The state of the dielectric still involves a superposition of these tracks, but their interaction with external patterns has the same general features as classical interactions, namely strength, long range and altogether a many-body behavior. As an example of external patterns, one may think of the charge distribution on the electrodes, the details of the electric field and some other properties of the electric circuit in the detector, to which many smaller patterns can be added (for instance a charge distribution in a dislocation inside an electrode...). A track generates a polarization in the dielectric, which influences the charges on the electrodes, these charges, together with a change in the local electric permittivity of the dielectric inside the track, influence the electric field, and so on. There is action and reaction everywhere.\par

These interactions can modify the quantum probabilities of the tracks. A simple model showing this significant property consists in introducing the Hamiltonian for the system consisting of the patterns and the tracks. There is certainly no observable distinguishing between tracks and commuting with this Hamiltonian, so that quantum transitions can occur between different track states and modify their respective quantum probabilities. A remarkable property of these probability variations is that they can act between distant tracks. This should be contrasted with the abstract situation when there are no patterns and only a mathematical electric field, the  tracks are superposed in the quantum state of the dielectric. There are never two tracks existing simultaneously (no state of two tracks and only states involving a superposition of them), so that they cannot interact, just like different components of a wave function cannot interact: and all of them evolve unitarily without exchange and with unchanged probabilities. In a real situation, a different behavior is due to the back-reaction of the patterns over the state of the dielectric, as shown in the following explicit model.

\section{Classicality breaking}
\hspace*{\parindent}
A crude model involves two systems $T$ (for track) and $C$ (for classical). $T$ has a two-dimensional Hilbert space with basis vectors $|j >$ ($j = 1$ or 2) representing two possible tracks. $C$ consists of any number of identical patterns with their set of collective observables denoted by $X$. The Hamiltonian is taken as 
\beq
\label{1e}
H = (1/2M)P^2 + V(X) + W\ ,
\eeq

\noi where $V$ does not depend on the track and $W$ represents the track-patterns interaction. It is not supposed diagonal in the basis $|\alpha >$ and it will be convenient to write it as $W =  \lambda \cdot \sigma$, where $\sigma$ denotes the three Pauli matrices in the Hilbert space of $T$ and $\lambda$ a set of three effective couplings, generally depending on $X$ and on the time $t$ (because the interaction between a track and a pattern depends on their distance, on their relative motion and the degree of evolution of the tracks). One can express the evolution of the system by the wave function
$$|\phi>\ = \phi_1 (x,t)|1>\ + \ \phi_2 (x,t)|2>$$
 
\noi and express the classicality of $C$ through the usual expression $\phi_j = A_j exp(iS_j /\tilde{h})$, where the amplitude $A_j$ varies slowly with $x$ and the phase $S_j$ varies rapidly. The real and imaginary parts of the Schr\"odinger equation become then
\bea
\label{2e}
A_j \left \{ \partial S_j/\partial t + H_0 (\nabla S_j, x)\right \} &=& - \Sigma_k \ {\rm Re} \left \{ (\lambda \cdot \sigma )_{jk} \exp \left [ i(S_k - S_j)/ \hbar\right ] \right \} A_k \nn \\
&&+ (\hbar^2 /2M) \Delta A_j \ ,
\eea

\bea
\label{3e}
\partial A_j /\partial t +\nabla A_j \cdot  (\nabla S_j/M) &=& - (1 / \hbar) \Sigma_k\  {\rm Im} \left \{ \lambda \cdot \sigma_{jk} \exp \left [ i(S_k - S_j)/ \hbar\right ] \right \} A_k \nn \\
&&- \hbar A_j \nabla^2 S_j/2M \ ,
\eea

\noi where $j$, $k = 1$ or 2, $H_0(P, X ) =  (1/2M)P^2 + V(X)$ is the Hamiltonian for the patterns when there is no coupling with the tracks. If the tracks are created at time zero in a state $c_1|1> + c_2|2>$ , one can express the prior existence of patterns through initial wave functions $\phi_j = c_j \phi$ in which $\phi$ represents the unique state of patterns at that time. The left-hand sides of Eqs. (2) and (3) express respectively the Hamilton-Jacobi equation for $S_j$ and a transport  of $A_j$ along the classical trajectories resulting from $S_j$. \par

The collective system behaves like a pointer when there is entanglement between the states of $T$ and $C$, {\it i.e.} when $W$ is diagonal and only the coupling $\lambda_z$, associated with $\sigma_z$, does not vanish. The two channels still evolve classically though with two different classical Hamiltonians, respectively given by $H_0(p, x ) \pm \lambda (x,t)$, and the first term in the right-hand side of Eq. (3) vanishes. The quantum probabilities $p_j$, which are the squares of the Hilbert norms of the two wave functions $\phi_j$, are conserved. When the patterns do not behave as pointers however, the first terms in the right-hand side of Eqs. (2) and (3) do not vanish and oscillate more and more rapidly.  As a consequence, the quantum probabilities $p_j$ undergo significant fluctuations.  \par
 
Another remarkable consequence follows from the fluctuating terms in Eqs. (2) and (3): classicality is broken in the dynamics of patterns. This breaking can be expected to be strong in view of the denominator $\hbar$ in the first term of the right-hand side in Eq. (3).

\section{Squared norms versus probabilities }
\hspace*{\parindent}
The quantum probabilities $p_j$ were defined up to now as the square of norms, but the oscillations of the first terms in the right-hand side of Eqs. (2) and (3) behave as random fluctuations when one considers real patterns and their complexity:  an increasing number of them participate in the chain of interactions when the germs grow, and there are other sources of irregularity, such as thermal fluctuations (due to the interactions of the patterns with their own environment).  This kind of randomness is much more similar to the standard uncontrollable causes of randomness in the foundation of classical probability calculus than to an intrinsic quantum randomness. \par

It originates in the oscillating terms in Eqs. (2-3), which lead to erratic fluctuations in the squared norms $p_j$. When the number of patterns, big or small, is very large, as one can expect, these fluctuations become presumably chaotic. The wave function evolves nevertheless in a deterministic way under the Schr\"odinger equation with a fixed Hamiltonian. A chaotic motion, or even a very erratic one, can be approximated by probability calculus, when the source of apparent randomness is inaccessible. One will therefore speak from now on of  ``squared norms'' rather than quantum probabilities, and one will reserve the name ``probabilities'' to the chaotic behavior of the squared norms $p_j$.

\section{Brownian reduction }
\hspace*{\parindent}
One thus arrives at the following situation: Various virtual coarse-grained tracks have been produced in various detectors. They belong altogether to a quantum superposition (including a component ``no track anywhere'' when one detector is missing. They interact strongly at a distance with outside patterns, which lose their classicality in a contagious way, this chain of breakings extending eventually far away. There is also a back-reaction on the squared norms of the various tracks, which fluctuate erratically. The main question is then: What will be the final issue of this process? \par

The answer has been given long ago and thoroughly investigated since then by Pearle [10], although in another framework, and it goes as follows: Let one call ``channel $j$'' a definite coarse-grained track. Its squared norm is the square of an amplitude (it can be also expressed in more detail in terms of projection operators and density matrices if necessary) and it suffers erratic fluctuations because of interactions with various nearby patterns, so that it is a time-depending quantity $p_j(t)$. Let one denote the fluctuations occurring during a short time interval $\delta t$ by $\delta p_j(t)$. Their distribution will be supposed Gaussian and one has the condition
\beq
\label{4e}
\sum\nolimits_j p_j = 1\  , \quad \hbox{so that } \quad     \sum\nolimits_j \delta p_j = 1\  .
\eeq

The Gaussian averages $<\delta p_j>$ are supposed equal to zero and correlations can be defined as  $A_{jk} = - <\delta p_j \delta p_k>/\delta t$ when $j \not= k$, from which Eq. (4) yields $<\delta p_j^2> = (\sum_k A_{jk})\delta t$. These coefficients depend on time because of the development of ionization and they also depend on the various quantities $p_l$, which can be denoted collectively by $p$. The previous model suggests that they can depend also on other parameters, according to the amount of loss in classicality among the neighboring patterns, but this point will be ignored. Considering the fluctuations as very small and rapid, the vector quantity $p = \{ p_j \}$ moves randomly under a Brownian motion. One now introduces a probability distribution $P(p, t)$ for the random quantities $p_j(t)$, whose initial values $p_j(0)$ are given by the state of primary germs, when the particle crossed a detector. The sum of these quantities for all the tracks inside one detector $D$ is given by $|c|^2$, if $c$ is the amplitude for the corresponding value of $z$, {\it i.e.} by Born's probability rule. Using the diffusion approximation for Brownian motion, one finds that the distribution $P(p, t)$ is governed by the Fokker-Plank equation [10]:
\beq
\label{5e}
\partial P(p,t)/\partial t = \sum_{jk} \left ( \partial /\partial p_j - \partial/ \partial p_k \right )^2 \left \{ A_{jk} (p,t) P(p,t)\right \} \ .
\eeq

This equation must be completed by initial conditions and boundary conditions. The initial condition is given by: 
\beq
\label{6e}
P(p, 0) = \prod_j \delta \left ( p_j - p_j(0) \right ) \ .
\eeq
 							
\noi The boundary conditions specify what happens when one coordinate $p_j$ vanishes and $p$ reaches the boundary of the domain in which probabilities are positive and make sense. One may guess the result by means of a physical argument. Consider a hypothetical case where the electric field was screened in one of the coarse-grained parallelepipeds in the detector. There would be no track there. If the screening is removed later, when the tracks elsewhere have already become macroscopic, no new track will be created since the necessary occasion was lost when the particle crossed $D$. One will assume by analogy that a squared norm $p_j$ remains forever zero after it vanished once. This intuitive condition can be expressed as the so-called absorptive boundary condition for the diffusion equation (5).\par

The evolution of the distribution $P(p,t)$ is then well defined and its asymptotic behavior is given by Pearle's theorem:
\beq
\label{7e}
P(p , \infty ) = \sum\nolimits_j p_j (0) \ \delta (p_j - 1) \ ,
\eeq

\noi which is fundamental for an understanding of reduction according to the present views. It does not matter that many different virtual tracks can be possible inside various detectors (or what will be the final spark in some Geiger counter). It does not matter that the events behave differently in these detectors, or that different patterns play a role during the selection of a specific final track in a specific detector and in the selection of a final detector. The delta functions in Eq. (7) mean that, finally, one of the tracks and only one of them, say $j$ will come out in the following sense: $p_j$ will be ultimately equal to 1 and all the other squared norms $p_k$, $k \not= j$, vanish. \par

 This means that every component for another track has vanished, so that the final state of the track-pattern system involves only this unique state $j$. This is therefore equivalent to wave function reduction, as expressed in the Copenhagen approach [14]. The factor $p_j(0)$ in front of the delta function means that this event will be realized with a chaotic probability $p_j(0)$ during a series of so-called identical measurements, because every new measurement starts from a different wave function of the particle-patterns system and every tiny change in the state of patterns yields a different chaotic motion. Accordingly, Born's probability rule is valid for such a series. Moreover, since the state of the incoming particle is still entangled with the state of the surviving track, a second measurement of $Z$ will give an identical result.\par

If this interpretation of reduction is correct, observation suggests that classicality is restored at the end of the process, but though there are theoretical indications in this direction, no proof of this important conjecture can be proposed now. Another important requirement would be an estimate of the rate of reduction, but  only two formal remarks about it can be mentioned. One of them relies on a known property of Brownian motion in the absorptive case: the average time for the evolving quantity $p(t)$ to reach a boundary value (where one of its elements $p_j$ is equal to 1 and the other ones vanish) is the inverse of the smallest eigenvalue of the operator in the right-hand side of Eq. (5). But to get an estimate of this eigenvalue, one should know an order of magnitude of the correlation coefficients $A_{jk}$. The rough model in Eq.~(2) and (3) could be used tentatively for this purpose, but one would then have to understand well the relevant patterns as they act in a realistic experiment, together with their interactions, so that a more ambitious and rigorous  program is obviously necessary.\\

\noi \underline{\it The density matrix}

One considered a wave function of the track-pattern system in Section 4, but this system is not in a pure state. One must therefore consider the density matrix $\rho$ for the collective coordinates of this system (the so-called reduced density matrix). Its eigenvalues and eigenvectors are given by
\beq
\label{8e}
\rho = \sum_{\alpha} \pi_{\alpha} | \varphi_{\alpha}>  < \varphi_{\alpha}| \ .
\eeq

\noi If the wave function $\varphi_{\alpha}$ induces a fluctuation $\delta p_{j\alpha}$ in channel $j$ during a small time interval $\delta t$ and the fluctuations for different values of the index $\alpha$ are uncorrelated, the standard deviation of the total change $\delta p_j$ in the probability $p_j$ is given by
\beq
\label{9e}
 (\Delta \delta p_j )^2 = \Sigma_{\alpha} \pi_{\alpha} (\Delta \delta p_{j\alpha})^2\approx \delta p^2  \ , 
\eeq

\noi where $\delta p$ stands for an order of magnitude of the fluctuations already encountered in Sections 3 and 5. The main results are therefore unchanged.

\section{Conclusions}
\hspace*{\parindent}
This work does not propose a full-fledged theory of reduction but only a conjecture, based on several simple remarks. Some possible loopholes remain in several places and a more careful examination is obviously needed. The questions are however so many and sometimes so difficult that one should rather consider this proposal as stating a program, whose basic idea asserts that reduction is due to a breaking and restoration of classicality. More generally, this proposal would amount to a wider statement of stability for classicality. It could also yield a new interpretation for the meaning of probabilities in quantum mechanics, with no intrinsic quantum randomness and only a chaotic behavior of many macroscopic patterns during a measurement. However, even if these considerations were shown theoretically consistent, they would be valid only if quantitative estimates were obtained and were found appropriate.

\end{document}